\documentclass[prl,preprintnumbers,
eqsecnum,floatfix,letterpaper,superscriptaddress,nofootinbib,showpacs]{revtex4}
\usepackage{color}
\usepackage[normalem]{ulem}
\usepackage{amsmath,amssymb,graphicx}
\usepackage{bm}
\usepackage{times}
\usepackage{microtype}
\usepackage{booktabs}
\usepackage{subfigure}
\usepackage[normalem]{ulem}
\usepackage[varg]{txfonts}
\usepackage[colorlinks, pdfborder={0 0 0}]{hyperref}
\definecolor{LinkColor}{rgb}{0.75, 0, 0}
\definecolor{CiteColor}{rgb}{0, 0.5, 0.5}
\definecolor{UrlColor}{rgb}{0, 0, 0.75}
\hypersetup{linkcolor=LinkColor}
\hypersetup{citecolor=CiteColor}
\hypersetup{urlcolor=UrlColor}
\allowdisplaybreaks
\newcommand{\ui}{\mathrm{i}}

\begin{document}
\newcommand{\red}{\color{red}}
\newcommand{\blue}{\color{blue}}
\title{Testing the Binary Black Hole Nature of a Compact Binary Coalescence}
\date{\today}
\author{N. V. Krishnendu}
\email{krishnendu@cmi.ac.in}
\affiliation{Chennai Mathematical Institute, Siruseri, 603103, India.}
\author{K. G. Arun}
\email{kgarun@cmi.ac.in}
\affiliation{Chennai Mathematical Institute,  Siruseri, 603103, India.}
\author{Chandra Kant Mishra}
\email{ckm@iitm.ac.in}
\affiliation{Indian Institute of Technology Madras, Chennai - 600036, India.}
\affiliation{ICTS-TIFR, Bengaluru (North) - 560089, India.}

\begin{abstract} We propose a novel method to test the binary black hole nature
of compact binaries detectable by gravitational wave (GW) interferometers and,
hence,  constrain the parameter space of other exotic compact objects. The
spirit of the test lies in the ``no-hair" conjecture for black holes where all
properties of a Kerr black hole are characterized by its mass and spin.  The
method relies on observationally measuring the quadrupole moments of the
compact binary constituents induced due to their spins. If the compact object
is a Kerr black hole (BH), its quadrupole moment is expressible solely in terms
of its mass and spin. Otherwise,  the quadrupole moment can depend on
additional parameters (such as the equation of state of the object). The higher
order spin effects in phase and amplitude of a gravitational waveform, which
explicitly contains the spin-induced quadrupole moments of compact objects,
hence uniquely encode the nature of the compact binary. Thus,  we argue that an
independent measurement of the spin-induced quadrupole moment of the compact
binaries from GW observations  can provide a unique way to distinguish binary
BH systems from binaries consisting of exotic compact objects.  \end{abstract}

\preprint{}
\maketitle

{\it Introduction}: With the twin detections of binary black hole mergers by
advanced LIGO interferometers~\cite{AdvLIGO}, black holes (BHs) are no longer
just elegant mathematical entities but a physical
reality~\cite{Discovery,GW151226,O1BBH}. Now we know that BHs do exist in
nature, and they can form a binary BH system and merge emitting  gravitational
waves (GWs) to form a single BH.  Analytical frameworks of post-Newtonian
theory (PN)~\cite{Bliving} and BH perturbation theory~\cite{TSLivRev03}
together with numerical relativity~\cite{Pretorius07Review}  have provided us a
theoretical platform to study and interpret the GW observations of compact
binary mergers.  Both the observed events, GW150914 and GW151226,  were found
to be consistent with a binary black hole merger with approximate total masses
of $65 M_{\odot}$ and $22 M_{\odot}$, respectively.  The strong evidence for
their binary black hole (BBH) nature comes from the following facts~\cite{TOG}:
{{1) Keplerian estimates of the orbital size are naturally explained by
invoking a binary BH system, 2) the observed ringdown waveform is consistent
with the least-damped quasinormal mode of a Kerr BH~\cite{Kerr63} (with the
inferred final mass and spin), and finally,  3) the reconstructed signal
matches excellently with the numerical relativity waveforms of a BBH
merger.}}\\

With planned upgrades towards operating advanced LIGO detectors at respective
design sensitivities and more detectors (such as advanced
Virgo~\cite{AdvVirgo}, KAGRA and LIGO-India~\cite{LIGO-India}) joining the
worldwide network of GW interferometers, many more such detections are likely
to happen in the future observation runs~\cite{Rates}. One of the important
questions  from a fundamental physics viewpoint, is whether we can confidently
distinguish the mergers of BBHs from that of binaries comprised of exotic
compact objects such as gravastars~\cite{Gravastars} and boson
stars~\cite{BosonStars}, which may mimic many features of a BBH merger (see,
also  Ref.\cite{Giudice2016} for a recent review on possible BH mimickers and
their GW signatures). \\

The definition of a Kerr BH is very closely tied with the the ``no-hair"
conjecture which says that all the properties of a Kerr BH are completely
described by its mass and spin. The quasinormal mode spectrum of a Kerr BH that
is formed, say, by the merger of two compact objects would,  hence,  be
completely characterized by the mass and spin of the remnant BH. This is a
topic that has been  studied in great detail over the past two decades.
References \cite{BHspect04,BCW05} studied the abilities of GW detectors to
carry out spectroscopy of a remnant compact object thereby testing its BH
nature.  The possibility of constraining specific BH mimicker models such as
boson stars using quasinormal mode spectrum observations has been discussed in
\cite{BC06,Macedo:2013}. If we have a stellar mass BH orbiting a supermassive
BH or an intermediate mass BH the dynamics of the stellar mass BH  (treated as
a test particle) would encode information about the multipole structure of the
central BH,  and, therefore constrain any possible deviations from the BH
nature~\cite{Ryan97,BrownEtalIMRI2012,Hughes2001}. While these methods are
restricted to studying the BH nature of the central compact object, the recent
proposals in Refs.~\cite{CardosoTidal2017,Maselli:2017,Noah2017} showed how the
measurement of the tidal Love number of a compact binary may be used to detect
exotic compact objects constituting a compact binary.\\

\begin{figure}[t] \centering
{\includegraphics[height=2.in,width=3.0in]{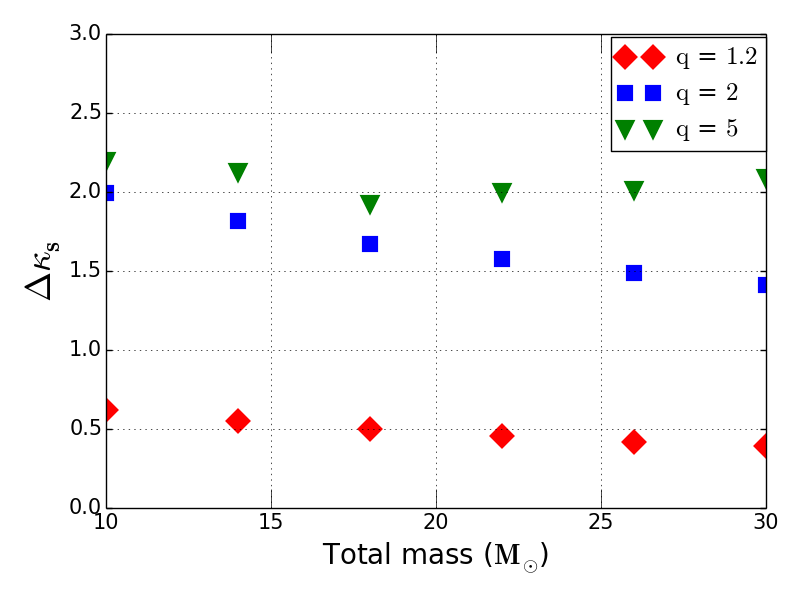}}
{\includegraphics[height=2.in,width=3.0in]{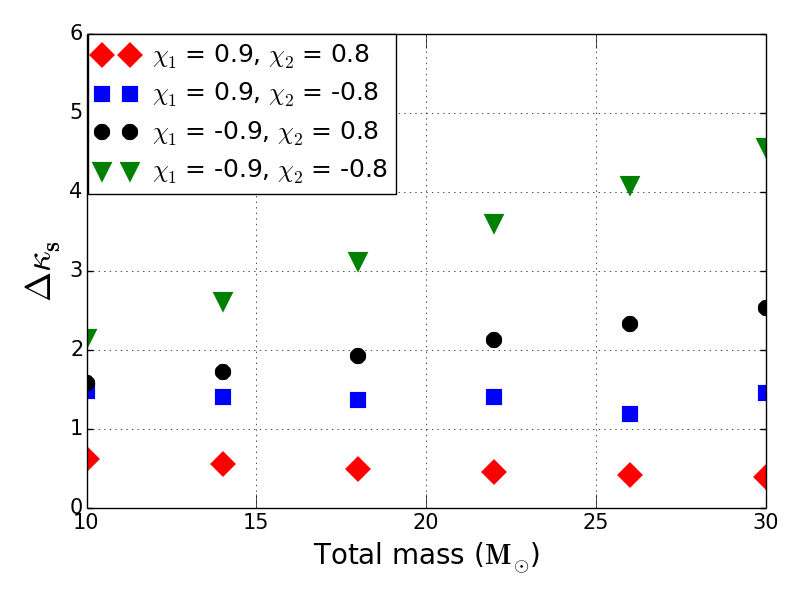}}
\caption{ Errors in measuring $\kappa_s$ as a function of the binary's total mass
for three different mass ratio cases (top panel) and for different spin
configurations (bottom panel) for advanced LIGO. The values of dimensionless
spin parameters ($\chi_1, \chi_2$) are fixed at 0.9 and 0.8 for the top panel
plots,  whereas mass ratio ($q$) is fixed to be 1.2 for the plots in the bottom
panel. Both panels assume a fixed inclination angle of the binary,
$\iota=\frac{\pi}{3}$. The binary's location and other angular parameters are chosen in a
way that produces an observed signal-to-noise ratio of 10.  }
\label{fig:summary} \end{figure}

In this Letter,  we propose a new method to test the binary black hole nature
of the detected GW event by measuring the spin-induced quadrupole moments of
the binary's constituents, whose values are unique for Kerr BHs in GR due to
the no-hair conjecture.  For an isolated Kerr BH, it is wellknown that the
quadrupole moment  scalar is given by $Q=-m^3\,\chi^2$, where $m$ is the mass
of the BH and $\chi$ is the magnitude of the dimensionless spin parameter
defined as ${\vec \chi}=\frac{{\vec S}} {m^2}$ (where ${\vec S}$ is the spin
angular momentum vector of the BH). For a non-BH compact object, this may be
generalized to $Q=-\kappa\; m^3\,\chi^2$, with $\kappa=1$ as the  BH limit.
Depending on the equation of state, studies have shown that for neutron stars
(NSs), $\kappa$ may range between  $\sim$2 and
14~\cite{Laar97,PappasMultipole2012}), for boson stars between $\sim$10 and
150~\cite{Ryan97b},  and for (thin shell) gravastars, $\kappa$ may even take
negative values~\cite{Uchikata2015} (which means the spin leads to prolateness
of the object instead of oblateness.)\\

In the PN model of compact binaries, the spin-induced quadrupole moment terms
appear at the same order where the leading order quadratic-in-spin terms appear
(note,  $Q\propto\chi^2$), which is second PN order~\cite{Poisson97}.  The
parameter  $\kappa$  that characterizes the magnitude of the spin-induced
quadrupole moment (given the nature of the object)  for each binary component
can be tagged as $\kappa_1$ and $\kappa_2$ following the notation of Ref.
\cite{Marsat2014} (throughout the paper, suffix 1 refers to the heavier compact
binary component and 2 the lighter one). If we rewrite the waveforms in terms
of the symmetric and antisymmetric combinations of $\kappa_1$ and $\kappa_2$
given by $\kappa_s=(\kappa_1+\kappa_2)/2$ and $\kappa_a=(\kappa_1-\kappa_2)/2$,
respectively, then a BBH system is specified by $\kappa_s=1, \kappa_a=0$. This
suggests, if we can accurately measure $\kappa_s$ and $\kappa_a$ to be 1 and 0,
respectively, we have established that the detected compact binary is a BBH.\\

However, note that $\kappa_s$ and $\kappa_a$ are highly degenerate parameters
whose simultaneous extraction turns out to yield almost no constraint on them
(this will have to be revisited using Bayesian methods in a future work).
Hence, we resort to a method where we fix  $\kappa_a$ to be 0, as expected for
a Kerr  BBH, and then calculate the error bars associated with the measurement
of $\kappa_s$ from GW observations. The aim here is to see how well can we
estimate $\kappa_s$ around the true value of 1 (for a BBH) and, hence,  confirm
that the observed system is indeed a BBH.  These error bars can be interpreted
as upper bounds on the value of $\kappa_s$  allowed for exotic compact objects.
In this sense, the proposed test is a ``null test" of the BBH nature, where,
observations would constrain the allowed range of deviations of $\kappa_s$ from
the BBH value. Moreover, since the spirit of the test relies on the fact that
quadrupole moments of BHs in a BBH system would depend only on the mass and the
spin, the proposed test can be regarded as the no hair theorem test for the
BBHs. \\

 We wish to clarify that the error bars  here refer  to the width of the
measured distribution of $\kappa_s$ at a fixed confidence level (in our case,
$1-\sigma$).  Depending on the masses and spins  of the system, this width may
be much larger than 1, in which case,  this may be better interpreted as an
upper bound on the allowed value of $\kappa_s$ for the given system. In most
cases we have studied  (in context of advanced LIGO), it is less than $\sim20$
(see Figs. 1 and 2).  Since $\kappa_s$ for interesting BH mimickers such as
boson stars can be as high as 150, the proposed method will be able to put
stringent, model-independent constraints on the parameter space of BH
mimickers. It should also be noted that  though we have posed this as a null
test, the proposed test can detect the signatures of exotic compact objects
through a shift in the peak of the measured distribution away from 1, as is
expected for BH mimickers.\\

In general, if we parametrize the deviation of $\kappa$ by $\kappa=1+\alpha$
(where $\alpha$ is the deformation parameter,  which is 0 for BHs)  and assume
that the constituents of the binary are of  identical types
($\alpha_1=\alpha_2$),  then, again, showing $\kappa_s=1$ is equivalent to
showing the BBH nature of the compact binary system. This is because we again
have $\kappa_a\equiv0$, which is consistent with our original assumption for
BBHs. Note that even if the detected compact binary constitutes  two stars
which have $\kappa \neq 1$, the proposed method will be sensitive in detecting
them as they will add to the systematic offset in the measured value of
$\kappa_s$ from 1.  Hence, our proposal to measure only $\kappa_s$ should work
for compact binaries with any combination of compact objects when applied to
the real data.\\

{\it Waveform model.} Because of  the recent progresses in the post-Newtonian
modeling of spinning compact
binaries~\cite{BFH2012,ABFO08,Marsat2014,BFMP2015,MKAF16}, we now have access
to the higher order spin corrections to the GW phasing and amplitude.  Here we
use a waveform which is 2PN in amplitude and  4PN (note that the phasing
formula at the 4PN only includes spin-orbit tail terms and hence is only
partial. See a related discussion in Ref.~\cite{MKAF16}) in phase and spins of
the two compact objects are considered to be along or opposite  the orbital
angular momentum vector of the binary. The spin-induced quadrupole moment
coefficient appears at 2PN, 3PN and 3.5PN orders.  The spin-induced octupole
moment coefficient which appear at 3.5PN is set to 1, the BH value as we focus
only on quadrupole here. See the section Supplemental Material below for
details of the waveform model.\\

{\it Estimation of $\kappa_s$:} We use the semianalytical parameter estimation
technique based on the Fisher information matrix formalism~\cite{CF94} to
deduce typical accuracies with which $\kappa_s$ may be estimated from GW
observations.  The Fisher information matrix approach allows us to calculate
the widths of the posterior distribution of various parameters for Gaussian
noise and in the limit of high signal-to noise-ratio (SNR) (see Ref.
\cite{Vallisneri07} for a detailed discussion on the possible caveats).  Unlike
previous works with PN waveforms which have subdominant modes (e.g. Refs.
\cite{ChrisAnand06b,AISSV07}), we truncate the waveforms at {\it twice} the
orbital frequency of the binary when it reaches the innermost stable circular
orbit ($2 F_{\rm ISCO}$) as opposed to the choice of $k F_{\rm ISCO}$, where
$k$ is the maximum number of harmonics of the orbital phase present in the
waveform. Here, the ISCO frequency is computed using numerical fitting formulas
listed in Eqs. (3.7) and (3.8) of Refs. \cite{HusaFit,EccPEFavata}. By doing
so,  we hope to control the systematics due to the neglect of merger and
ringdown.  Though much less realistic than numerical methods based on
algorithms such as Markov chain Monte Carlo (MCMC) calculations, the
semianalytic method used here is significantly inexpensive in terms of
computational time and is expected to match  the predictions of numerical
methods in the high SNR limit~\cite{BalDhu98}.  However, we caution that the
errors we quote here should be taken as a typical order of magnitude of the
expected errors which will be quantified in the future with MCMC
investigations.\\

\begin{figure}[htp] \centering
{\includegraphics[height=2.in,width=3.0in]{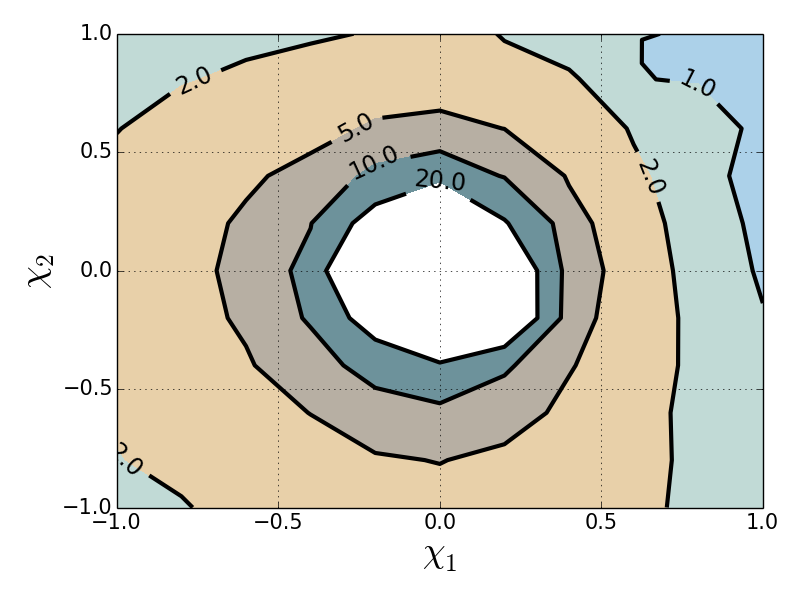}}
{\includegraphics[height=2.in,width=3.0in]{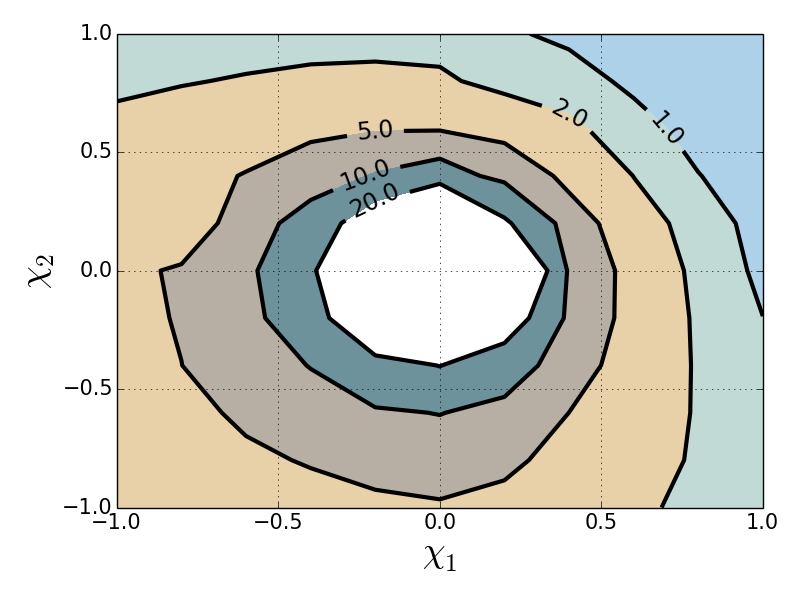}}
\caption{ Two-dimensional error contours indicating the measurability of
$\kappa_s$ in the $\chi_1-\chi_2$ plane for two representative binary systems:
$(5, 4)M_{\odot}$ (top panel) and $(10, 9)M_{\odot}$ (bottom panel) for
advanced LIGO sensitivity. The inclination angle of the binary is chosen to a
value of $\pi/3$, and the source is located and oriented in such a way that it
produces a signal-to-noise ratio of 10 at the detector.
}\label{fig:SpinCP}\end{figure}

For every system of interest, we construct a Fisher information matrix using
the waveform model discussed above for the set of parameters $\{t_c, \phi_c,
D_L,\iota, {\cal M}, \delta, \chi_1, \chi_2, \kappa_s\}$ which describe the
signal.  Here, $t_c$ and $\phi_c$ denote time and phase of the waveform at
coalescence, two mass parameters ${\cal M}={(m_1 m_2)^{3/5}/(m_1+m_2)^{2/5}}$
and $\delta={|m_1- m_2|/(m_1+m_2)}$  are known as the chirp mass and difference
mass ratio of the binary, parameters $(\chi_1, \chi_2)$ denote the
dimensionless spins of the binary components, and finally, $D_L$ and $\iota$
are the luminosity distance and the inclination angle of the binary,
respectively.  We consider the problem from a single detector standpoint and,
hence,  do not include the angles which describe the source location in the set
of parameters.  We compute the lower bound on the errors of each parameter
(Cramer-Rao bound) by taking the square root of the diagonal values of the
inverse of the $9\times9$ Fisher information matrix (covariance matrix). These
errors are calculated for different masses and spins of the compact binary
systems as well as for different inclination angles ($\iota$). We consider the
sources to be located and oriented in such a way that they produce a SNR of 10
at the detector. Projected advanced LIGO noise PSD~\cite{Ajith2011} is used to
compute the errors. The $1-\sigma$ error bars on $\kappa_s$ (with a peak at 1)
assume $\kappa_a=0$, which is the case for Kerr BBHs. From a  GW event, if we
find that the posterior distribution for $\kappa_s$ is offset from 1, it may be
taken as a signature for at least one of the binary components to be a non-BH
object.  Throughout the Letter, we quote errors in the measurement of
parameters characterizing the spin-induced effects. However, as mentioned
earlier, for many parts of the parameter space, we find that errors are larger
than 100\% for which the quoted errors should be considered as  "bounds'" on
the parameter in question.\\

{\it Results and discussion.} The dependences of the errors (for a fixed SNR of
10) in measuring $\kappa_s$ as a function of the total mass, for few mass ratio
cases (top panel) and spin configurations for a near-equal mass system (bottom
panel) for advanced LIGO sensitivity are shown in Fig.~\ref{fig:summary}. This
clearly shows that the proposed test works very well for {\it highly spinning,
near-equal mass systems}.  Evidently, the observed improvement for rapidly
spinning systems can be attributed to the large spin-induced quadrupole moment
they possess.  In addition, for nearly equal mass systems, the best estimates
of $\kappa_s$ come from compact binaries in which the spins of both components
are aligned with respect to the orbital angular momentum vector of the binary,
and the worst estimates are for those cases where the component spins are
antialigned with respect to the  orbital angular momentum.  The decrease in the
errors with mass ratio may be attributed to the additional mass ratio and
inclination angle dependences that amplitude corrections bring in, which affect
the correlation of $\kappa_s$ with other parameters (especially spins)  in a
nontrivial way leading to the observed trend. On the other hand, the dependence
of the errors on the spin orientation is due to its effects on the upper cutoff
frequency.  The figure shows that even with a moderate SNR of 10, the proposed
test works very well for a number of mass ratio and spin configurations, where
the best cases have $\Delta\kappa_s<0.5$ (50\%). It is worth recalling that the
allowed values of $\kappa_s$ for BBH mimickers, such as binaries involving
boson stars, can be as high as 150.  Hence, the expected bounds are capable of
putting stringent constraints on those models. \\

Figure~\ref{fig:SpinCP} displays the dependence of the errors of $\kappa_s$ on
the component spins for two representative stellar mass compact binaries with
component masses $(5, 4) M_{\odot}$ and $(10, 9)M_{\odot}$. The results are
very promising and show that for dimensionless spins larger than 0.5, the
errors in estimating $\kappa_s$ are smaller than $\sim5$ in both the cases.
This would mean that the proposed test could be effective in certain cases even
with moderate spins.\\

Since the GW detectors are poised to observe tens to hundreds of BBH mergers in
the coming years, we also have the interesting possibility of combining the
constraints from these individual observations.  If there are $N$ detections,
the errors go down by roughly a factor of $\sqrt{N}$. Hence, the combined
posterior of about 100 events on the null hypothesis may narrow down the
constraints on $\kappa_s$ by a factor of 10.\\

\begin{figure}[htp] \centering
{\includegraphics[height=2.in,width=3.0in]{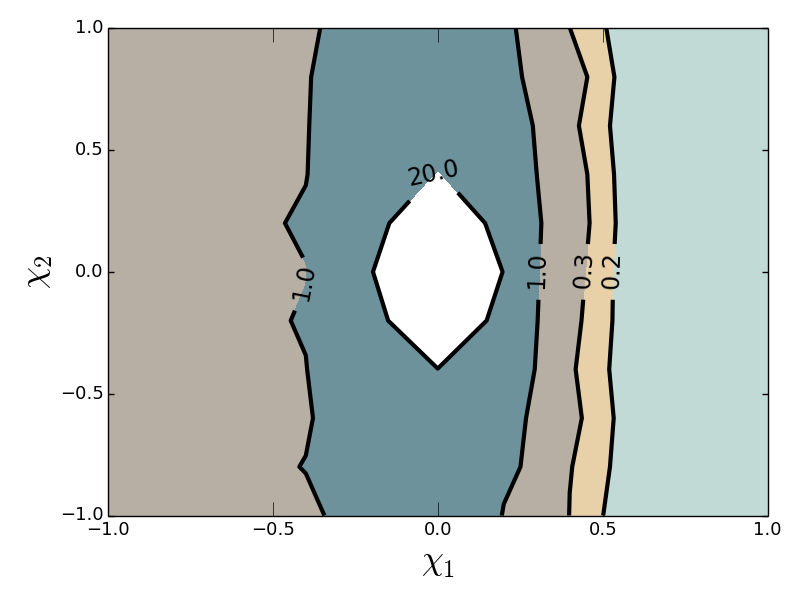}}
{\includegraphics[height=2.in,width=3.0in]{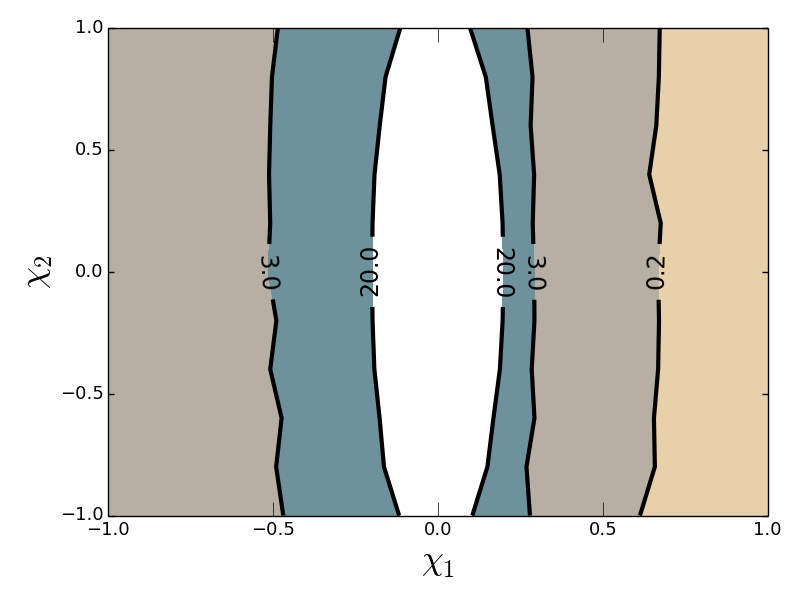}} \caption{
Projected constraints from GW observations of SMBBH mergers
by the LISA detector as a function of the component spins for two representative
SMBBH configurations  $(5 \times 10^{6}, 10^{6})M_{\odot}$ (top panel) and
$(10^{7}, 10^{6})M_{\odot}$ (bottom panel) located at 3 Gpc. The inclination
angle of the binary is chosen to a value of $\pi/3$.  }\label{fig:eLISA1}
\end{figure}

{\it Possible constraints on $\kappa_s$ from space-based detectors.} With the
recent success of of LISA pathfinder mission~\cite{Pathfinder}, there is
renewed interest in pursuing a GW detector in space with low frequency
sensitivity capable of observing supermassive BBH (SMBBH) mergers.  Towards
this goal, we extend our study to the case of low frequency space-based
detectors like LISA and projected constraints possible on $\kappa_s$ from them.
The results are shown in Fig.~\ref{fig:eLISA1} which uses the noise PSD of
Ref.~\cite{Babak2017}.  The SMBBH system is assumed to be at a luminosity
distance of 3 Gpc. We find that the LISA observations of SMBBH  mergers can
very accurately constrain the $\kappa_s$ parameter and,  hence, confirm the BBH
nature of the observed sources, tightly constraining any alternatives to BBHs.
It should be clear from Fig.~\ref{fig:eLISA1} that errors in measuring
$\kappa_s$ are smaller than 10\% for a number of configurations with moderate
spins, making the test an extremely deep probe of any possible deviation from
BBH nature.  These results show how LISA can be a very sensitive probe of
fundamental physics.\\

{\it  Possible constraints on BH mimickers.} Since boson stars can have
$\kappa$ between $\sim10$ and $150$~\cite{Ryan97b}, binary systems of boson
stars may have $\kappa_s$ in the range $\sim10-150$.  This allowed range lies
well within the reach of the proposed test.  Recently, for slowly rotating thin
shell gravastars, Ref.~\cite{Uchikata2015} showed that the spin-induced
quadrupole can take a wide range of values depending on the specifics of the
model (see Fig. 7 of Ref.  ~\cite{Uchikata2015}).  This range includes
$\kappa=1$, the BH value, too.  Indeed, if $\kappa_{\rm GS}=1$, our test will
not be able to distinguish it from a BH. Except for this very fine-tuned
scenario, the projected bounds from the proposed test might significantly help
to constrain the allowed parameter space of gravastars and can influence the
theoretical developments in the field. The details of the bounds possible on
specific BH mimicker models will be reported elsewhere~\cite{KAM2016b}.\\

We note that the proposed test may not be very sensitive in distinguishing a BH
from a BH mimicker in a NSBH system. This is because the neutron stars are
expected to have small spins ($\leqslant0.05$) for which spin-induced
quadrupole would be very small. Moreover, since NSs are expected to have
$\kappa $  value in the range of $2-14$, very accurate estimation of the
$\kappa$ parameters of both the binary components is necessary to make the
above distinction.  This may be possible only with the future generation of GW
detectors.\\

There are some effects which can potentially contaminate the effectiveness of
the proposed test. Because the compact objects in binaries are, strictly
speaking, not isolated, the no-hair conjecture holds only approximately due to
which there can be systematic effects which may affect the test  (see
Ref.~\cite{Campanelli08} for a discussion on this aspect). Further, if the BHs
are charged, then the resulting values of $\kappa$ will be offset from the Kerr
value. Lastly, the choice of upper cutoff frequency may be different from ours
if the object has structure and, hence, can cause systematic errors in our
estimates. These issues need more careful examination which will be carried out
in the future.\\

We conclude by noting that  once implemented in a Bayesian framework, this
proposal can be used to represent every detected compact binary system as
contours in the $\kappa_1-\kappa_2$ space.  Using multiple observations, the
joint posteriors can tighten the bounds from this proposed null test,
potentially constraining the parameter space allowed for non-BH compact
objects.  Inclusion of precessional features in the waveform and incorporating
this effect into effective one body waveforms or phenomenological waveforms,
which capture merger and ringdown phases as well, are likely to yield tighter
constraints and will be explored in the future.\\

{\it Acknowledgement.} K. G. A. and N. V. K.  were partially supported by a
grant from Infosys Foundation.  K. G. A. acknowledges the support by the
Indo-US Science and Technology Forum through the Indo-US Centre for the
Exploration of Extreme Gravity (Grant No. IUSSTF/JC-029/2016). This work was
initiated following a suggestion by Luc Blanchet. We thank  N. J.-McDaniel for
very useful discussions on many aspects, especially those related to boson
stars and gravastars. We thank Walter Del Pozzo for useful comments on the
manuscript. We are thankful to P. Ajith, G.  Faye,   B. R. Iyer, S.  Kastha,
A. Laddha, and B. S. Sathayprakash for very useful discussions. This research
has significantly benefitted from the interactions during the "Future of
Gravitational Wave Astronomy Workshop" at the International Centre for
Theoretical Sciences (Code No.  ICTS/Prog-fgwa/2016/04).  This document has
LIGO preprint number {\tt P1600340}.

\section {Supplemental Material} \subsection{Spin-induced quadrupole and
octupole pieces in compact binary waveforms}

The waveform used in this work is a variant of the one that is presented in
Ref.~\cite{MKAF16}. These are constructed by simply making the dependences on
parameters characterising the spin-induced quadrupole moment (through
$\kappa_s$ and $\kappa_a$) and spin-induced octupole moment (through
$\lambda_s$ and $\lambda_a$) explicit in the waveform, which were set to their
respective values for Kerr BHs while writing the waveform model of
Ref.~\cite{MKAF16}. In this note we list various pieces of the waveform where
such dependences occur.\\ 

Let us first recall the schematic expression for the frequency domain amplitude
of a gravitational wave signal, $\tilde h(f)$, given in Ref.~\cite{MKAF16}.
\footnote{Pre-factor of Eq.~1 of Ref.~\cite{MKAF16} should be multiplied with a
factor ${1/\sqrt \eta}$. We have corrected this in the Eq.~\eqref{eq:hf}.} This
reads

\begin{equation} \tilde{h}(f)=\frac{M^2}{D_L}
\sqrt{\frac{5\,\pi}{48\eta}}\sum_{n=0}^{4}\sum_{k=1}^{6}
V_k^{n-7/2}\,C_{k}^{(n)}\,e^{\ui (k\,\Psi_\mathrm{SPA}(f/k)-\pi/4)} \, .
\label{eq:hf} \end{equation}

Here, $M$, $\eta$ and $D_L$ denote the total mass, symmetric mass ratio
parameter and the distance to the binary, respectively and the indices $n$ and
$k$ denote the PN order and harmonic number, respectively. The coefficients
${\cal C}_{k}^{(n)}$ denote the amplitude corrections associated with the
contribution from $k{\rm th}$ harmonic at $n{\rm th}$ PN order. Related
expressions for each of the ${\mathcal C}_{k}^{(n)}$s can be found in
Ref.~\cite{ABFO08, MKAF16}. Here we list the only coefficient which has
explicit dependence on the parameters ($\kappa_s$ and $\kappa_a$) and
corresponds to the contributions from the $2$nd harmonic at the 2PN order
(${\cal C}_{2}^{(4)}$). In addition, $\Psi_{\rm SPA}$ represents the phase of
the first harmonic in the frequency domain as obtained under the Stationary
Phase Approximation (SPA) (see sec. VI of Ref.~\cite{ABFO08} for details on
SPA). Schematically the expression for this phase can be written as follows

\begin{equation} \Psi_\mathrm{SPA}(f) = 2\pi f t_\mathrm{c} - \phi_\mathrm{c} +
\left\{\frac{3}{128\eta\,v^{5}}
\left[\psi_\mathrm{NS}+\psi_\mathrm{SO}+\psi_\mathrm{SS}
+\psi_\mathrm{SSS}\right]\right\}_{v=V_1(f)},  \end{equation}

where $\phi_\mathrm{c}$ denotes the orbital phase at the instant $t_\mathrm{c}$
of coalescence. \\

Further, one can write the spin part of the SPA phase more explicitly as 

\begin{equation} \psi_\text{Spin}\equiv\psi_\mathrm{SO}+\psi_\mathrm{SS}
+\psi_\mathrm{SSS}
={v^3}\left[\mathcal{P}_3+\mathcal{P}_4\,v+\mathcal{P}_5\,v^2
+\mathcal{P}_6\,v^{3}+\mathcal{P}_7 v^4+\mathcal{P}_8 v^{5} +\cdots\right].
\label{eq:psi_so} \end{equation}

Again expressions for the coefficients $\mathcal{P}_n$ can be found in
Ref.~\cite{ABFO08, MKAF16} where explicit dependence on $\kappa_s$ and
$\kappa_a$ is suppressed by setting them to their respective values for Kerr
BBHs. Here we provide expressions for coefficients that contain explicit
dependence on $\kappa_s$ and $\kappa_a$. Below we list the amplitude/phase
coefficients that do contain explicit dependence on $\kappa_s$ and $\kappa_a$
and can be combined to those listed in Ref.~\cite{ABFO08, MKAF16} to write the
final waveform expression. 

\begin{eqnarray} \mathcal{C}_2^{(4)}&=& \frac{1}{\sqrt{2}}
\Biggl\{F_+\Bigg[\frac{113419241}{40642560}+\frac{152987}{16128}\,\eta
-\frac{11099}{1152}\,\eta^2 +\left(\frac{165194153}{40642560}
-\frac{149}{1792}\,\eta+\frac{6709}{1152}\,\eta ^2\right)c_{\iota}^2 
+\left(\frac{1693}{2016}-\frac{5723}{2016}\,\eta
+\frac{13}{12}\,\eta^2\right)c_{\iota}^4
\nonumber\\&-\frac{1}{24}&\left(1-5\,\eta+5\,\eta ^2\right) c_{\iota}^6
+(1+c_{\iota}^2)\biggl[ (\boldsymbol{\chi}_\mathrm{s}\cdot
\hat{\boldsymbol{L}}_\mathrm{N})^2 \left(\frac{1}{32}+\frac{23 \,\eta
}{8}+\frac{3 \delta  \kappa _a}{2}+\frac{3}{2}\left(1-2 \,\eta \right) \kappa
_s\right)
+(\boldsymbol{\chi}_\mathrm{a}\cdot \hat{\boldsymbol{L}}_\mathrm{N})^2
\left(\frac{1}{32}-3 \,\eta +\frac{3 \delta  \kappa _a}{2}
\right.\nonumber\\&+&\left. \frac{3}{2}\left(1-2 \,\eta \right) \kappa_s\right)
+(\boldsymbol{\chi}_\mathrm{a}\cdot
\hat{\boldsymbol{L}}_\mathrm{N})(\boldsymbol{\chi}_\mathrm{s} \cdot
\hat{\boldsymbol{L}}_\mathrm{N}) \left(\frac{\delta }{16}+3\,(1-2 \,\eta )
\kappa _a+3 \delta  \kappa _s\right) \biggr]\Bigg]
+\ui\,c_{\iota}\,F_{\times}\Bigg[\frac{114020009}{20321280}
+\frac{133411}{8064}\,\eta-\frac{7499}{576}\,\eta^2 \nonumber \\ &+&
\left(\boldsymbol{\chi}_\mathrm{s}\cdot
\hat{\boldsymbol{L}}_\mathrm{N}\right){}^2 \left(\frac{1}{16}+\frac{23}{4}\eta
+3 \delta  \kappa _a+3\,(1-2 \,\eta ) \kappa _s\right)+
\left(\boldsymbol{\chi}_\mathrm{a}\cdot
\hat{\boldsymbol{L}}_\mathrm{N}\right){}^2 \left(\frac{1}{16}-6 \,\eta +3
\delta  \kappa _a+3\,(1-2 \,\eta ) \kappa _s\right)
+(\boldsymbol{\chi}_\mathrm{a}\cdot
\hat{\boldsymbol{L}}_\mathrm{N})(\boldsymbol{\chi}_\mathrm{s}\cdot
\hat{\boldsymbol{L}}_\mathrm{N})\left(\frac{\delta }{8}
\right.\nonumber\\&+&\left.6\,(1-2 \,\eta ) \kappa _a+6 \delta  \kappa
_s\right)
+\left(\frac{5777}{2520}-\frac{5555}{504}\,\eta+\frac{34}{3}\,\eta^2\right)c_{\iota}^2
-\frac{1}{4}\left(1-5\,\eta+5\,\eta^2\right) c_{\iota}^4\Bigg]\Biggr\}\,
\Theta(2 F_\mathrm{cut} - f) \label{eq:C24} \end{eqnarray}

\begin{subequations} \begin{eqnarray} \label{eq:P4} {\mathcal P}_4 &=&
-\frac{5}{8}(\boldsymbol{\chi}_\mathrm{s}\cdot\hat{\boldsymbol{L}}_\mathrm{N})^2\,
\Bigl[1+156 \,\eta +80 \,\delta  \,\kappa _a+80 (1-2 \,\eta ) \kappa _s\Bigr]
+(\boldsymbol{\chi}_\mathrm{a}\cdot\hat{\boldsymbol{L}}_\mathrm{N})^2
\left[-\frac{5}{8}-50 \,\delta  \,\kappa _a-50 \kappa _s+100 \,\eta
\left(1+\kappa _s\right)\right] \nonumber\\
&-&\frac{5}{4}(\boldsymbol{\chi}_\mathrm{a}\cdot\hat{\boldsymbol{L}}_\mathrm{N})(\boldsymbol{\chi}_\mathrm{s}\cdot\hat{\boldsymbol{L}}_\mathrm{N})
\Bigl[\delta +80\,(1-2 \,\eta )\,\kappa _a+80 \,\delta  \,\kappa _s\Bigr]\,,\\
{\mathcal P}_6 &=& \pi\,\Biggl[\frac{2270}{3}\,\delta
\,\boldsymbol{\chi}_\mathrm{a}\cdot\hat{\boldsymbol{L}}_\mathrm{N}
+\left(\frac{2270}{3}-520\,\eta\right)
\boldsymbol{\chi}_\mathrm{s}\cdot\hat{\boldsymbol{L}}_\mathrm{N}\Biggr]
+(\boldsymbol{\chi}_\mathrm{s}\cdot\hat{\boldsymbol{L}}_\mathrm{N})^2
\left[-\frac{1344475}{2016}+\frac{829705}{504}\,\eta+\frac{3415}{9}\,\eta ^2
+\delta \left(\frac{26015}{28}-\frac{1495}{6}\,\eta\right) \kappa_a
\right.\nonumber\\&+&\left.  \left(\frac{26015}{28}-\frac{44255}{21} \,\eta
-240 \,\eta ^2\right) \kappa _s\right]
+(\boldsymbol{\chi}_\mathrm{a}\cdot\hat{\boldsymbol{L}}_\mathrm{N})^2
\left[-\frac{1344475}{2016}+\frac{267815}{252} \,\eta -240 \,\eta ^2+\delta
\left(\frac{26015}{28} -\frac{1495}{6} \,\eta \right) \kappa
_a+\left(\frac{26015}{28} \right.\right.\nonumber\\&-&\left.\left.
\frac{44255}{21} \,\eta -240 \,\eta ^2\right) \kappa _s\right]
+(\boldsymbol{\chi}_\mathrm{a}\cdot\hat{\boldsymbol{L}}_\mathrm{N})(\boldsymbol{\chi}_\mathrm{s}\cdot\hat{\boldsymbol{L}}_\mathrm{N})
\left[\left(\frac{26015}{14}-\frac{88510}{21} \,\eta -480 \,\eta ^2\right)
\kappa _a+\delta\, \biggl[-\frac{1344475}{1008}+\frac{745}{18} \,\eta
+\left(\frac{26015}{14} \right.\right.\nonumber\\&-&\left.\left.  \frac{1495
}{3} \,\eta\right) \kappa _s\biggr]\right]\,, \label{eq:P6} \\
{\mathcal P}_7 &=& \delta\,\boldsymbol{\chi}_\mathrm{a}\cdot
\hat{\boldsymbol{L}}_\mathrm{N}\left(-\frac{25150083775 }{3048192}
+\frac{26804935}{6048}\,\eta-\frac{1985}{48}\,\eta^2\right)
+\boldsymbol{\chi}_\mathrm{s}\cdot\hat{\boldsymbol{L}}_\mathrm{N}\left(-\frac{25150083775}{3048192}+\frac{10566655595}{762048}\,\eta
-\frac{1042165}{3024}\,\eta^2
\right.\nonumber\\&+&\left.\frac{5345}{36}\,\eta^3\right)+
(\boldsymbol{\chi}_\mathrm{s}\cdot\hat{\boldsymbol{L}}_\mathrm{N})^3\,
\Biggl[\frac{265}{24}+\frac{4035 \,\eta }{2}-\frac{20 \,\eta
^2}{3}+\left(\frac{3110}{3}-\frac{10250 }{3}\,\eta +40 \,\eta ^2\right) \kappa
_s+\delta \,\biggl[\left(\frac{3110}{3}-\frac{4030 \,\eta }{3}\right) \kappa _a
\nonumber\\&-&440\,(1-\eta) \lambda _a\biggr] -440\,(1-3 \,\eta ) \lambda
_s\Biggr] +(\boldsymbol{\chi}_\mathrm{a}\cdot\hat{\boldsymbol{L}}_\mathrm{N})^3
\Biggl[\left(\frac{3110}{3}-\frac{8470}{3} \,\eta \right) \kappa _a-440\,(1-3
\,\eta ) \lambda _a+\delta\,\biggl[\frac{265}{24}-2070 \,\eta
\nonumber\\&+&\left(\frac{3110}{3}-750 \,\eta \right) \kappa _s -440(1-\eta )
\lambda _s\biggr]\Biggr]
+(\boldsymbol{\chi}_\mathrm{s}\cdot\hat{\boldsymbol{L}}_\mathrm{N})^2
(\boldsymbol{\chi}_\mathrm{a}\cdot\hat{\boldsymbol{L}}_\mathrm{N})
\Biggl[\left(3110-\frac{28970}{3} \,\eta +80 \,\eta ^2\right) \kappa
_a-1320\,(1-3 \,\eta ) \lambda _a \nonumber\\&+&\delta
\biggl[\frac{265}{8}+\frac{12055 }{6}\,\eta +\left(3110-\frac{10310
}{3}\,\eta\right) \kappa _s-1320\,(1-\eta) \lambda _s\biggr]\Biggr]
+(\boldsymbol{\chi}_\mathrm{a}\cdot\hat{\boldsymbol{L}}_\mathrm{N})^2
(\boldsymbol{\chi}_\mathrm{s}\cdot\hat{\boldsymbol{L}}_\mathrm{N})
\Biggl[\frac{265}{8}-\frac{6500}{3} \,\eta +40 \,\eta ^2
\nonumber\\&+&\left(3110-\frac{27190 \,\eta }{3}+40 \,\eta ^2\right) \kappa _s
+\delta  \biggl[\left(3110-\frac{8530 \,\eta }{3}\right) \kappa
_a-1320\,(1-\eta ) \lambda _a\biggr]-1320\,(1-3 \,\eta ) \lambda _s\Biggr] \,.
\label{eq:P4} \end{eqnarray} \end{subequations}

\bibliographystyle{apsrev} 
\bibliography{ref-kappa.bib}

\end{document}